\documentclass[%
 reprint,
superscriptaddress,
 amsmath,amssymb,
 aps,
 pra,
]{revtex4-2}
\usepackage[english]{babel} 
\usepackage{graphicx}
\usepackage{dcolumn}
\usepackage{bm}

\usepackage{enumitem}
\usepackage{titlesec}
\usepackage{diagbox}
\usepackage{etoolbox}
\usepackage{booktabs}
\usepackage{tabularx}
\usepackage{makecell}
\usepackage{hyperref}
\usepackage[dvipsnames]{xcolor}
\usepackage[normalem]{ulem}

\makeatletter
\renewcommand{\section}{%
  \@startsection{section}{1}{\z@}%
    {-3.5ex \@plus -1ex \@minus -.2ex}%
    {2.3ex \@plus.2ex}%
    {\normalfont\bfseries\scshape\centering}%
}
\makeatother

\titlespacing*{\subsection}
{0pt}{1.5ex plus 1ex minus .2ex}{1ex plus .2ex}
\titlespacing*{\subsubsection}
{0pt}{1.5ex plus 1ex minus .2ex}{1ex plus .2ex}
\titlespacing*{\paragraph}
  {\parindent}      
  {\medskipamount}  
  {1.5em}           

\usepackage{graphicx}
\usepackage{array, multirow}
\usepackage{physics}
\usepackage{ragged2e}

\usepackage{caption}
\usepackage{subcaption}
\DeclareCaptionFormat{justifcaption}{%
\begin{minipage}{\linewidth}\noindent\justifying\small #1#2#3\end{minipage}}
\DeclareCaptionFormat{centercaption}{%
\begin{minipage}{\linewidth}\noindent\Centering\small #1#2#3\end{minipage}}
\usepackage{tikz}
\usetikzlibrary{positioning,arrows.meta,calc}
\usepackage{standalone}

\usepackage{accents}

\makeatletter
\renewcommand{\fnum@figure}{\textbf{Fig.~\thefigure}}  

\makeatother

\def\arxiversion{}

\ifdefined\arxiversion
    \newcommand{\heading}[1]{\section{#1}}
    \newcommand{\subheading}[1]{\subsection{#1}}
    
    \newcommand{\arxiv}[1]{#1}
    \newcommand{\natcomm}[1]{}
\else 
    \newcommand{\heading}[1]{\section*{#1}}
    \newcommand{\subheading}[1]{\subsection*{#1}}

    \newcommand{\arxiv}[1]{}
    \newcommand{\natcomm}[1]{#1}
    
    \makeatletter
    \patchcmd{\@startsection}
    {\@hangfrom{\csname #1name\endcsname \@seccntformat{#1}}}
    {\@hangfrom{\csname #1name\endcsname \@seccntformat{#1}}}
    {}{} 

    \renewcommand{\@seccntformat}[1]{%
      \ifcsname the#1\endcsname
       \csname the#1\endcsname\quad 
    \fi
    }
    \makeatother

    \preto{\subsection}{\setcounter{paragraph}{0}} 
\fi

\begin{document}

\title{Real-time BP decoding of a high-rate QEC code}

\author{Andreas Kell}
    \email{Andreas.Kell@d-fine.com}
\affiliation{d-fine GmbH}

\author{Daniel Herr}
    \email{Daniel.Herr@d-fine.com}
\affiliation{d-fine AG}

\author{Daniel Bascones García}
    \email{danibasc@ucm.es}
\affiliation{Complutense University of Madrid}

\author{Alexandru Paler}
    \email{alexandru.paler@aalto.fi}
\affiliation{Aalto University, School of Science, Department of Computer Science, Konemiehentie~2, 02150 Espoo, Finland}
\affiliation{PlanQC GmbH, M\"unchener Str.~34, 85748 Garching, Germany}

\author{Davide Orsucci}
    \email{davide.orsucci@dlr.de}
\affiliation{DLR - Institute of Communications and Navigation}

\author{Francisco Lázaro}
    \email{francisco.lazaroblasco@dlr.de}
\affiliation{DLR - Institute of Communications and Navigation}

\author{Francisco García Herrero}
    \email{francg18@ucm.es}
\affiliation{Complutense University of Madrid}

\begin{abstract}
We present a real-time FPGA decoder for a high-rate $[[512,174,8]]$ 3D product qLDPC code. We construct a circuit-level detector-error model for a compact syndrome-extraction schedule and propose a small parallel ensemble of row-wise serial min-sum decoders with cumulative memory. The decoder matches $\text{BP+OSD-CS}_\text{10}$ accuracy with one core and Relay-BP accuracy with three cores. 
The FPGA implementation meets the latency budget of neutral-atom and trapped-ion systems. An ASIC implementation would improve the latency, enabling real time decoding even for superconducting systems.
At a physical error rate of $p=10^{-3}$, three FPGA devices decode 174 logical qubits with a logical error rate of $5\times10^{-7}$ and effective distance 6.
For $p=10^{-4}$, and assuming the number of BP iterations is capped to 30, the decoder's average latency is $5.1\,$\textmu s and its worst-case latency is $97.9\,$\textmu s.
\end{abstract}

\maketitle

High-rate quantum error correction (QEC) codes reduce the number of physical qubits, two-qubit gates, and classical decoding operations needed per logical qubit. They can also simplify the control stack, because a larger fraction of the physical qubits directly contributes to logical information. This is especially relevant for quantum-memory experiments, where control overhead and classical post-processing can be major design drivers~\cite{fellous_optimizing_2023}. The bivariate bicycle (BB) family has been a standard choice for accurate decoders, but its rates are modest: distance-6 codes have rates near 17\%, while the gross code $[[144, 12, 12]]$ has a rate of only 8.3\%~\cite{bravyi_high-threshold_2024}. Newer high-rate constructions, including Kasai, clustered-cyclic, and mitten codes, improve this trade-off by encoding more logical qubits at a given block length while retaining competitive logical error rates~\cite{Komoto_Quantum_2025,gu_qgpuparallellogicquantum_2026,bhardwaj2026highrateqldpcprocessors}. 
However, these higher rate codes generally need more complex decoders. For example, the Kasai-code implementation in~\cite{zhao_towards_2026} uses belief propagation (BP), followed by Relay-BP, and mixed integer programming decoder to decode distance-12 codes with up  to 580 logical qubits and a rate above 50\%, but CPU latencies are in the millisecond range, and optimistic estimates for FPGA/GPU implementations reduce the latency to
approximately $700\,$\textmu s~\cite{kschischang_factorgraph_2001,muller_improvedbeliefpropagationsufficient_2025,gu_scalableneuraldecoderspractical_2026}. 
Clustered-cyclic codes can reach accuracy comparable to the gross BB code at a higher rate and support directly addressable logical measurements, but they require ordered statistics decoding (OSD). A full OSD order-0 implementation needs roughly $2N+3M$ cycles, where $N$ and $M$ are the numbers of error nodes and detectors, and filtered versions reduce latency at the cost of accuracy~\cite{Derks_designing_2025,Panteleev_degenerate_2021,Roffe_decodingacross_2020,bascones_exploring_2025,maurya_fpgatailoredalgorithmsrealtimedecoding_2026}. 
Furthermore, the latency introduced by FPGA communication between BP and OSD has, as far as we know, not been studied.
Mitten codes encode 108 logical qubits at 20\% rate and distance 18, but decoding these codes results in worst-case estimated FPGA latencies exceeding $400\,$\textmu s~\cite{bhardwaj2026highrateqldpcprocessors}. 

Other existing FPGA-based implementations for high-rate codes have focused either on graphs corresponding to the code capacity regime, such as \cite{valls_syndromebased_2021} or \cite{crest_layereddecodingquantumldpc_2023}, or on phenomenological noise models \cite{raveendran2023soft}, which are considerably simpler than a more complete circuit-level noise model. Circuit-level noise model proposals have required multiple FPGAs to process approximately 12 logical qubits \cite{maurer_realtimedecodinggrosscode_2025}, treating X and Z errors as independent, or have required ensemble sizes involving nearly thirty cores within almost ten FPGA devices to adequately handle correlated errors \cite{bascones_scalablefpgaarchitecturerealtime_2026}.

\begin{figure}
    \centering
    \includegraphics[width=0.48\linewidth]{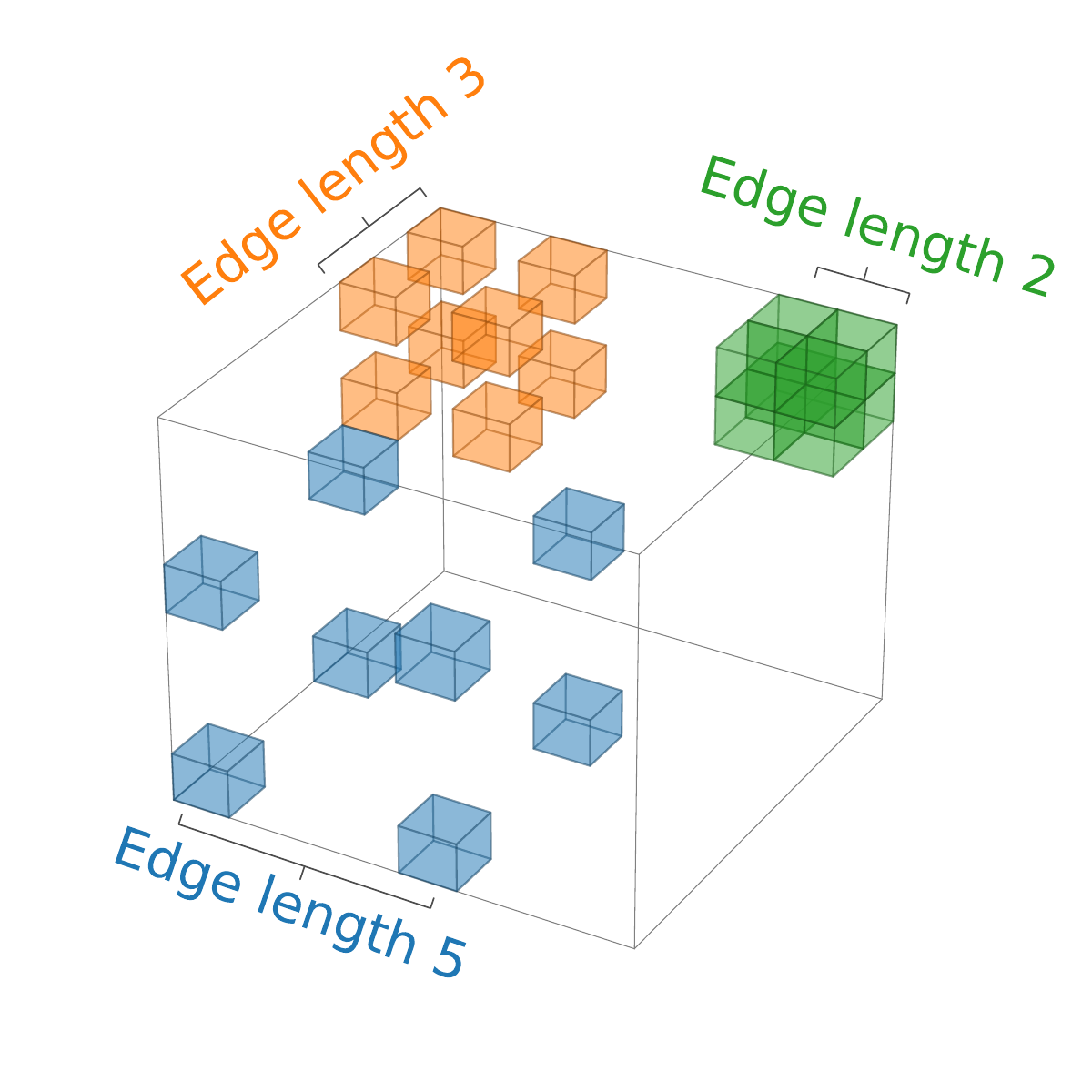}
    \includegraphics[width=0.48\linewidth]{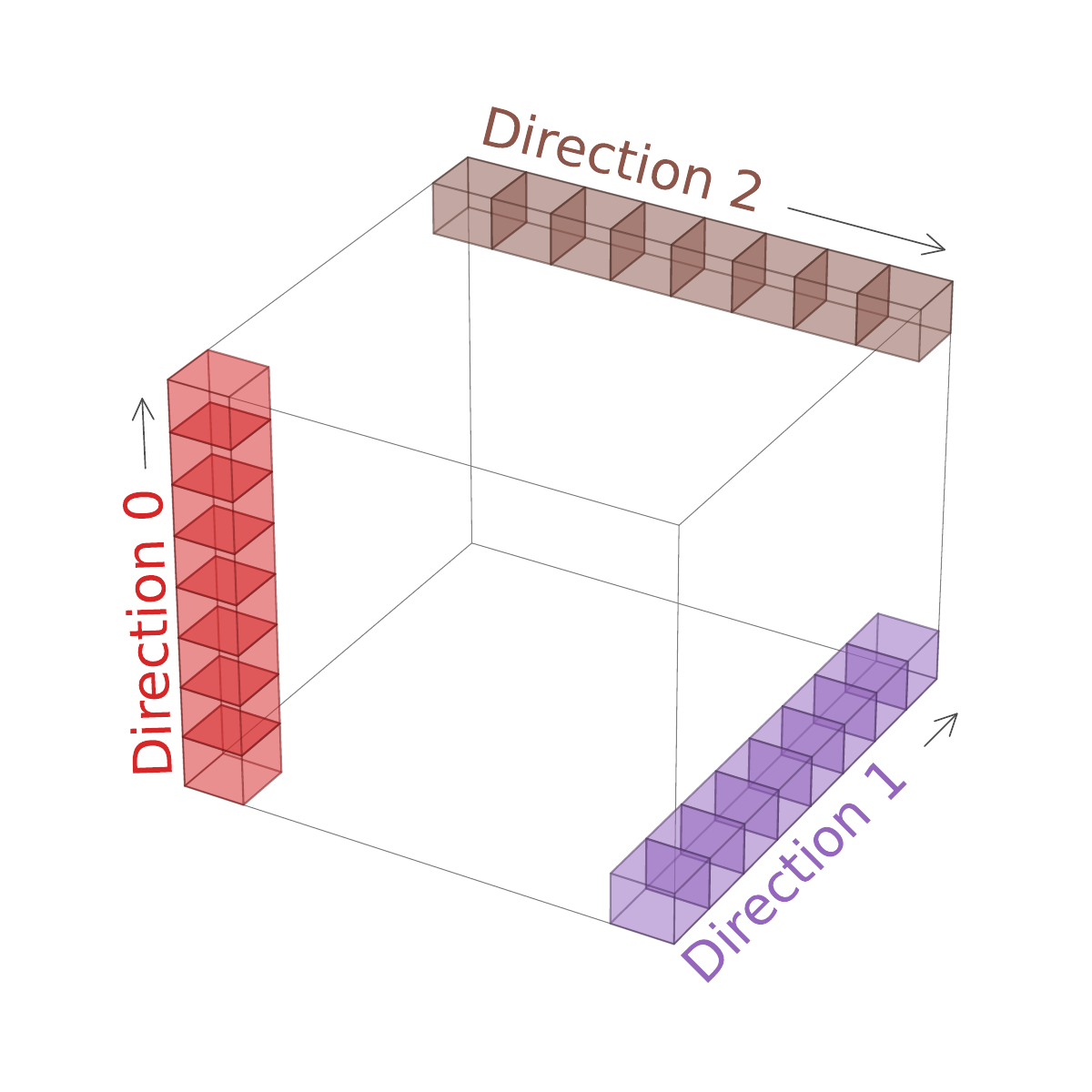}
    \caption{\textbf{Three-dimensional view of selected stabilizers.} Left: cube-like $Z$ stabilizers with edge lengths 2, 3, and 5. Right: row-like $X$ stabilizers in the three coordinate directions of the $[[512,174,8]]$ code.}
    \label{fig:3DCube}
\end{figure}

Our work bridges this gap by presenting a low-latency decoder for a high-rate code.
In particular, we consider the $[[512,174,8]]$ 3D product code from~\cite{ostrev_classical_2024}, whose rate is 33.98\%. Its stabilizers are arranged on a three-dimensional lattice: cube-like $Z$ stabilizers and row-like $X$ stabilizers, as shown in Fig.~\ref{fig:3DCube}. This structure is attractive for neutral-atom systems because most (though not all) stabilizers involve neighboring qubits. 
The main contribution of this work is a real-time decoding architecture for this code under circuit-level noise. We first construct a detector-error model (DEM) for a compact syndrome-extraction circuit. We then implement a decoder that uses only a modified memory-BP variant~\cite{Chen_improvedbp_2025}: a row-wise serial min-sum decoder with cumulative memory, run in a small parallel ensemble. This ensemble does not run in relay fashion and does not require OSD or integer programming. 
Our decoder achieves a logical error rate per logical qubit of $5\times10^{-7}$ at a physical error rate of $10^{-3}$.
The decoder core runs at $3.264\,$\textmu s per iteration on an AMD xcvu19p‑fsva3824‑2‑e FPGA \cite{amd_vu19p}.
At a physical error rate $p=10^{-4}$, a single decoder core achieves an average latency of $5.1\,$\textmu s.
Thus, a contemporary ASIC could reach sub‑µs latencies \cite{bascones_exploring_2025,Barber_realtime_2025}, matching the requirements of superconducting architectures \cite{googleai_suppressing_2023}.  
For neutral‑atom or trapped‑ion platforms \cite{bluvstein_fault_2026,tripier_fault-tolerant_2026}, temporal multiplexing on the same FPGA would process more than 250 batches of 174 logical qubits within a 1\,ms budget, i.e., more than 40\,k logical qubits using only three devices.  
By contrast, CPU/GPU systems can meet the latency for a single 174‑qubit code but lack multiplexing without adding many cores and complex I/O.

\heading{Decoding Accuracy}

\begin{figure}
    \includegraphics[width=\columnwidth]{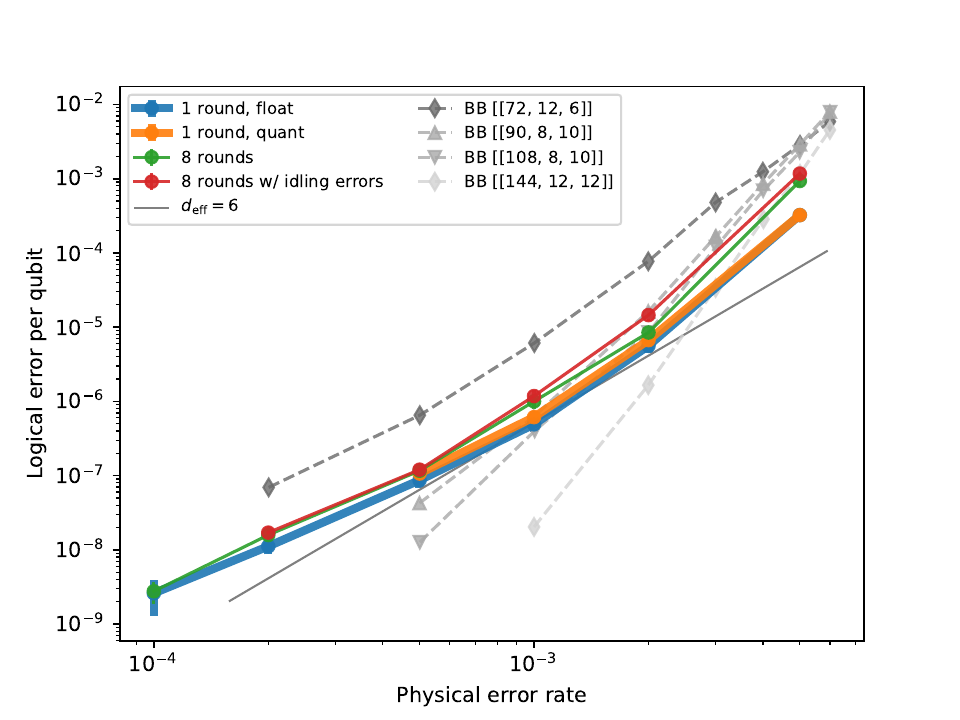}
    \caption{
    \textbf{Logical error rate per logical qubit.}
    Shown are the per-round results for eight rounds of syndrome extraction and one noiseless round of the 3-D product code with and without idling errors, obtained with a floating point decoder.
    Also shown are the results for a single round of syndrome extraction without idling errors followed by a final $Z$‑basis measurement of the 3‑D product code, obtained with a floating point and a quantized decoder.
    A ruler is provided which indicates the exponent $\left\lfloor d_\text{eff} / 2 \right\rfloor$ corresponding to the effective distance $d_{\text{eff}} = 6$.
    For comparison, the per‑logical-qubit, per‑round results of the BB codes decoded with $\text{BP+OSD-CS}_\text{10}$
    from~\cite{bravyi_high-threshold_2024} are plotted.
    } 
    \label{fig:performance}
\end{figure}

Figure~\ref{fig:performance} shows the logical error rate per logical qubit and per round. We use eight syndrome-extraction rounds, corresponding to the nominal distance, followed by one noiseless $Z$-basis measurement. 
For a physical error rate in a range around $10^{-3}$, we observe an effective distance $d_\text{eff}=6$, as the logical error rate behaves as $p_L \propto p^{\left\lfloor d_{\text{eff}} / 2\right\rfloor}$.
The achieved accuracy is competitive to BB codes of comparable distance. In particular, the logical error rate per qubit is below that of the [[72, 12, 6]] BB-code, which has the same effective distance.
Additionally, as expected from the lower distance of the 3-D product code, the achieved accuracy is lower compared to the gross code, BB [[144, 12, 12]]

The figure also shows that introducing idling errors has only a small effect. 
The reason for this is, as will be discussed later, that there are only idling errors at the very beginning and at the very end of the experiment, and no new idling errors are introduced when the number of syndrome extraction rounds is increased.

For the FPGA implementation, we use the simpler one-round setup without idling errors, ending in a noisy $Z$-basis measurement. Fixed-point quantization does not significantly degrade performance, as shown in Fig.~\ref{fig:performance}.

\heading{Decoding Latency}

\begin{figure}
    \includegraphics[width=.48\linewidth]{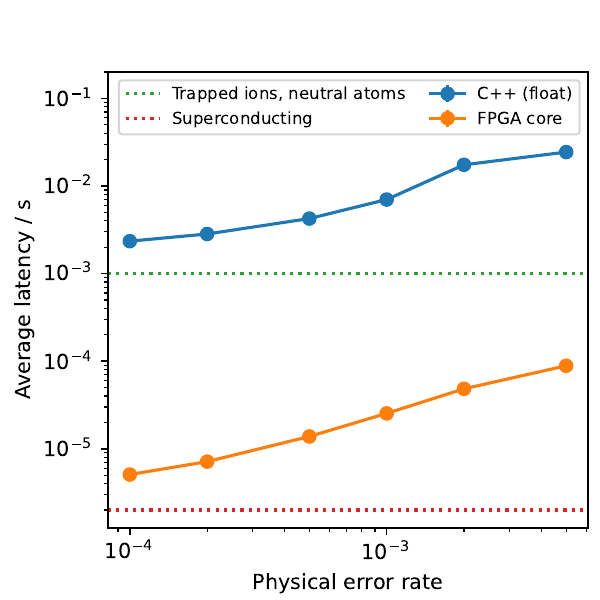}
    \includegraphics[width=.48\linewidth]{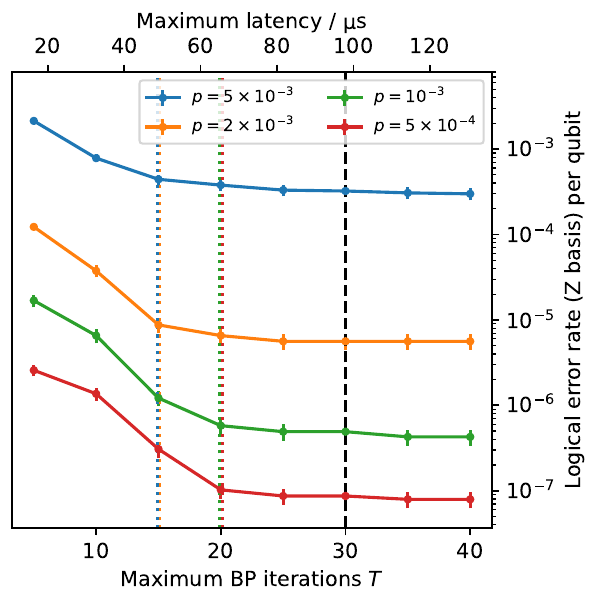} 
    \caption{
        \textbf{Average and maximum decoder latency.}
        \textbf{Left:} Shown are the average latency of our C++ decoder and of one
        implemented FPGA decoder core in comparison with the latency threshold of neutral-atom, trapped-ion and superconducting systems.
        \textbf{right:} shown is the logical error rate of the ensemble as a function of the maximum number of BP iterations $T$ (lower X-axis) or maximum latency of the FPGA core (upper X-axis) for various physical error rates $p$. The logical error rate decreases significantly only up to a certain maximum number of iterations $T$ or maximum latency (dotted vertical lines) and we choose $T=30$ as a default (black dashed vertical line).
    }
    \label{fig:latency}
\end{figure}

The latency of the FPGA implementation is analyzed in Fig.~\ref{fig:latency}.
As shown on the right side, the maximum number of iterations can be restricted to 30, which implies a worst-case latency for
the decoder core of $97.9\,$\textmu s.
This leaves sufficient margin for the integration into the control setup of a neutral-atom or trapped-ion system for real-time error correction, assuming a maximum latency of about 1 ms \cite{bluvstein_fault_2026, tripier_fault-tolerant_2026}.

The average latency, which is relevant when addressing the backlog problem, is displayed on the left side of Fig.~\ref{fig:latency}.
The queues associated with the time distribution are also relevant to the backlog problem. However, in the proposed approach, they are limited to 97.9~\textmu s when the BP decoder reaches the maximum number of iterations, making them significantly shorter than those associated with other post-processors, such as OSD.
For the lowest considered physical error rate of $10^{-4}$, the average latency per decoding is $5.1\,$\textmu s for the FPGA core, as the average number of iterations is 1.6, so that an ASIC implementation could realistically reach the required average latency.
Note, however, that the syndrome measurements of the 3D product code requires long range interactions which, at the time of writing, are difficult to realize in superconducting platforms.

\heading{Conclusions}


This work demonstrates that a high-rate 3D product code can be decoded in real time on a small number of FPGAs under circuit-level noise. The main contribution is a hardware-friendly decoding scheme that reaches the accuracy of more complex BP+OSD and Relay-BP schemes without their additional stages. The code encodes 174 logical qubits at a 33.98\% rate relying on just three FPGAs. The measured FPGA core latency is in the microseconds, with a bounded worst case. For $p=10^{-3}$, the logical error rate of $5\times10^{-7}$ is competitive with BB codes of similar effective distance, although it does not reach the accuracy of the higher-distance gross code.

The practical significance is that the same decoder hardware can be time-multiplexed over many logical-qubit sets, an option not easily available to CPU- or GPU-based solutions without a much larger core count.
The latency of the FPGA implementation is also more deterministic than that of a CPU implementation, which is important when many decoder instances must meet a common deadline.

Several issues are left to future work. We have not analyzed the feasibility of the syndrome-extraction circuit on superconducting platforms, nor the integration into a real control system, including communication links, quantization-area-latency trade-offs, or folded architectures for faster early termination. We also envision multi-FPGA integration of the three cores, ASIC implementation, and extension to a sliding-window decoder for multi-round syndrome extraction.

\heading{Methods}


\subheading{Circuit}
The syndrome-extraction circuits are built with \textit{Stim}~\cite{gidney_stim_2021}. For a single-round experiment, the circuit contains one round of $X$ and $Z$ stabilizer measurements and a final measurement of all data qubits. Each single-qubit gate, reset, and input and two-qubit gate receives depolarizing noise, and each measurement can flip with probability $p$. Depolarizing noise for idling qubits is only considered when explicitly stated. Hook errors up to weight 4 are possible for weight-8 stabilizers~\cite{manes_distance-preserving_2025}. We mitigate them by reordering CNOT gates within stabilizers \cite{tomita_low-distance_2014} and by interleaving $X$ and $Z$ stabilizer measurements \cite{tremblay_constant-overhead_2022, bravyi_high-threshold_2024}. In a multi-round experiment, the same circuit without its final measurement can be repeated with shifted cycles so that, except for the unavoidable beginning and end of the experiment, qubits are not left idle.

\begin{figure}
    \includegraphics[width=\columnwidth]{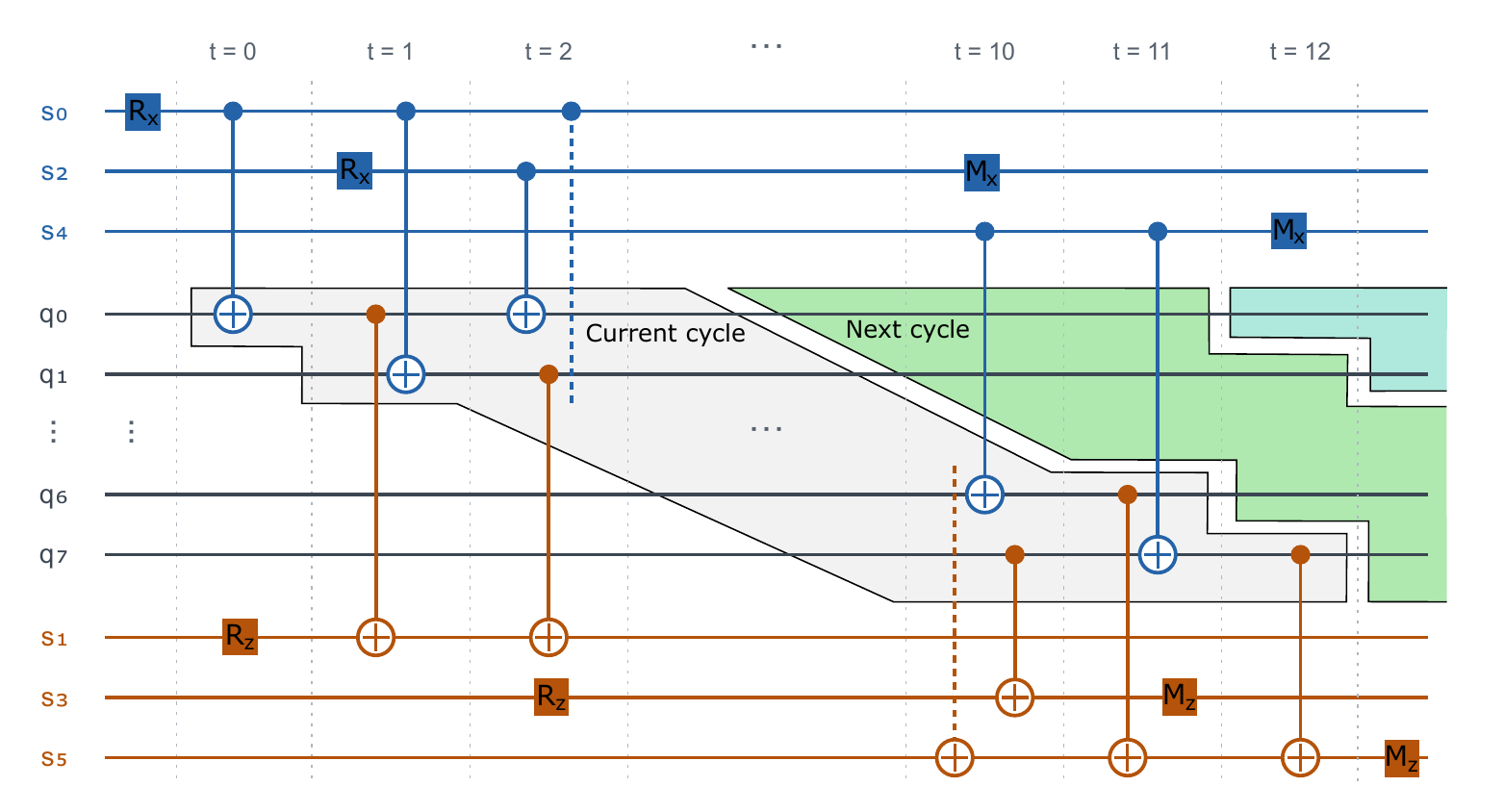} 
    \caption{
        \textbf{Sketch of syndrome extraction circuit} The circuit for one representative X or Z stabilizer per group (blue or red lines) and one representative data qubit per group (black lines) is displayed.
        The gray, green and turquoise polygons indicate how subsequent cycles can be arranged in a compact way in a multi-round setup.
    }
    \label{fig:schedule}
\end{figure}

The measurement schedule, presented in Fig.~\ref{fig:schedule}, is constructed as follows.
The stabilizers are partitioned into numbered groups $s \in {0, \dots}$ so that no two stabilizers in the same group measure the same data qubit.
The data qubits are also partitioned into numbered groups $q \in {0, \dots}$ so that no two data qubits in the same group are measured by the same stabilizer.
Then the schedule is defined such that a stabilizer in group $s$ measures a data qubit in group $q$ at time step $t=s + q$.
This guarantees that there is no collision, i.e., same qubit measured multiple times or same stabilizer applied multiple times in the same time step,
and that the commutation relations of the stabilizers are fulfilled because the stabilizer groups are applied in the same order for each data qubit.
By alternately choosing X and Z stabilizers for the groups, an interleaved schedule is created.
The numbering of the qubits defines the order of CNOT gates within a stabilizer.

This construction is similar to left-right circuits \cite{strikis_high-performance_2026}. It generalizes the partition of qubits to multiple subsets, instead of the two subsets left and right.
The coloring of Tanner graph edges is simplified by instead partitioning the stabilizers.

For this code, the six natural stabilizer groups are: $X$ stabilizers in directions 0, 1, and 2, and $Z$ stabilizers with edge lengths 2, 3, and 5, as shown in Fig.~\ref{fig:3DCube}. A valid data-qubit grouping is analogous to filling a Sudoku grid: each data-qubit number must be unique within each $X$ stabilizer row and $Z$ stabilizer block. We use
\begin{equation*}
\begin{aligned}
q(x,y,z)
={}&
\bigl(
      x_0 + 2x_1
    + y_2 + 2y_0
    + z_1 + 2z_2
\bigr) \bmod 4
\\
&{}+
4\bigl(
    \pi(x) \oplus \pi(y) \oplus \pi(z)
\bigr),
\label{eq:q-direct}
\end{aligned}
\end{equation*}
with the binary digits $x_0,x_1,x_2$ such that $x=x_0+2x_1+2^2x_2$, and similarly for $y$ and $z$, and with the parity bit $\pi(x)=x_0\oplus x_1\oplus x_2$. This assignment uses XOR in the high bit, which gives a better hook-error distribution than a simple modulo construction.

The single-round schedule spans 13 time steps because some qubits idle at the beginning and end. For repeated rounds, however, the next cycle can begin after the first group of CNOTs finishes at $t=5$, giving an effective depth of six time steps. No extra reset or measurement steps are required if a fresh ancilla set is used every second cycle.  Reusing ancillas would add two steps.

\begin{figure}
    \includegraphics[width=\columnwidth]{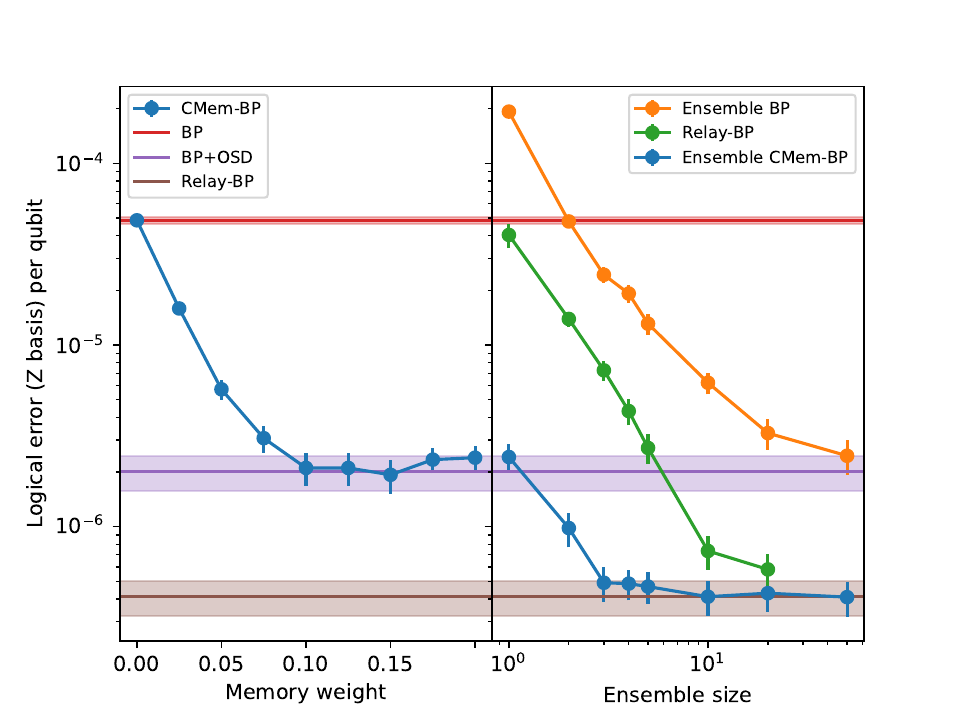}
    \caption{
        \textbf{Optimization of the decoder.}
        \textbf{Left:} The logical error rate of a single CMem-BP decoder is shown against the memory weight.
        \textbf{Right:} The logical error rate for an ensemble of CMem-BP decoders is shown against the ensemble size.
        For comparison, the results for an ensemble of BP without memory and for a Relay-BP decoder are displayed. For the latter, the number of legs is interpreted as ensemble size and all solutions are considered (i.e., $S=R$).
        \textbf{Both:} As a reference, the results for BP without memory, $\text{BP+OSD-CS}_\text{10}$ and Relay-BP-5 with 601 legs are given.
    }
    \label{fig:optimize_decoder}
\end{figure}

\subheading{Decoder}
For decoding, several variants of belief-propagation (BP) have been considered.
Row-wise serial updating combined with memory terms has proven particularly effective.
The memory-based update of the bias terms is set to 
\begin{equation}
    \Lambda_j(k) = \Lambda_j(k-1) + \gamma \left(M_j(k-1) - \Lambda_j(0)\right),\label{eq:memory-update}
\end{equation}
with memory weight $\gamma$ and the marginal
\begin{equation}
M_j(k) = \Lambda_j(k) + \sum_{i' \in \mathop{N}(j)}{\mu_{i'\rightarrow j}(k)}
\end{equation}
and the check-node to value-node messages $\mu_{i'\rightarrow j}$.
Note that standard Mem-BP is recovered if the first term of Equation~\ref{eq:memory-update}, $\Lambda_j(k-1)$ is replaced by $\Lambda_j(0)$, i.e., the messages are effectively integrated.
Therefore, we call this variant cumulative memory BP, CMem-BP, to distinguish it from other memory-based variants. CMem-BP matches the performance of $\text{BP+OSD-CS}_\text{10}$, as shown in Fig.~\ref{fig:optimize_decoder}.

The resulting logical error rate can be further decreased by employing an ensemble of multiple BP decoders.
Each decoder of the ensemble uses a different random processing order of check nodes.
Additionally, the a priori probability of each error node is set to zero with a probability of $5\,\%$.
After processing, the ensemble member that best matches the measured syndrome is selected by (i) choosing the solution whose syndrome has the smallest Hamming distance to the observed syndrome (zero distance indicates convergence) and (ii) picking, among those, the one with the lowest total a‑priori error probability.
This attains Relay‑BP‑level performance using only a small ensemble of three decoders, as shown in Fig.~\ref{fig:optimize_decoder}.

\subheading{FPGA Implementation} 
Given the stringent timing, area, and power constraints, a memory-based min-sum decoder was chosen over a sum-product decoder, whose additional adders and comparators in the check-node update would push the critical path beyond the target clock rate. Flooding was discarded because it does not converge reliably without a post-processing stage or a relay of roughly twenty legs (Fig.~\ref{fig:optimize_decoder}). A fully parallel message-passing mapping was also infeasible: the DEM would require routing hundreds of thousands of signals simultaneously among processing units, preventing a reasonable operating frequency.

Serial schedules therefore provide the best accuracy-to-hardware trade-off. Row-wise (horizontal) serial decoding with memory matches the accuracy of min-sum plus OSD of order 10, and its logical error rate is essentially identical to column-wise (vertical) decoding. We choose row-wise because it offers far more parallelism within each iteration: the DEM has 576 rows (check nodes) but 20{,}080 columns (error nodes), a factor of 34.9. Column-wise decoding would visit one error node after another sequentially, making it impractical at the required clock rate.

The implemented decoder uses normalized min-sum with a normalization constant of 0.625, which provides a good trade-off between latency, area, power, and accuracy. A memory term derived from the a priori probabilities in the DEM is stored in BRAM for the entire decoding process and applied at every iteration as defined in Eq.~\eqref{eq:memory-update} with $\gamma=0.125$, chosen for good decoding accuracy and simple implementation with bit shifts instead of multipliers. The priors, check-to-variable messages, and biases are quantized to fixed-point format: 5 word bits (0 fractional) for priors, 11 word bits (4 fractional) for messages, and 6 word bits (3 fractional) for biases. This quantization does not significantly degrade decoding performance (Fig.~\ref{fig:performance}). As shown in Fig.~\ref{fig:optimize_decoder}, three such cores are needed to form the ensemble that reaches the reported accuracy.

A single decoder core implements one check-node processing unit (CNU) and 397 variable-node processing units (VNUs) in parallel. During warm-up, the soft a priori values from the DEM are written into dedicated BRAM blocks. In the first iteration these values initialize the variable-to-check messages; in all subsequent iterations the same stored values provide the memory term. 
The columns are permuted via a graph-colouring step \cite{tarable_mapping_2004} so that, within any row, the up to 397 message addresses map to distinct BRAM banks, eliminating intra-row read/write conflicts.

Inside each VNU, the extrinsic variable-to-check message is computed by subtracting the previous iteration's check-to-variable message (held in a per-message buffer) from the current soft data and adding the memory weight from the prior BRAM. All arithmetic is performed in full precision; saturation and flooring are then applied to bring the result back into the selected fixed-point range. Detector values are fed into the core through an $M$-to-$1$ parallel-to-serial converter to avoid adding a multi-cycle load phase.

The CNU output messages are stored in a buffer of constant width equal to the maximum check-node degree. To reduce the CNU critical path and achieve a latency of one cycle per check node, $n_p$ pipeline stages are inserted, each adding one extra address to the message buffers. The row order of the parity-check matrix is chosen so that any two check nodes sharing a variable node are separated by at least $n_p$ rows, eliminating read-after-write hazards. The same permutation is applied consistently to the syndrome-bit presentation. Stalling the pipeline to avoid conflicts was considered but would increase the per-check-node latency by a factor of $(1+n_p)$, which is not acceptable under the timing budget.

After the pipeline, each pair of (original input, CNU output) is combined with a full-precision adder to produce the updated variable-node message, which is quantized and written to a separate message BRAM. In parallel with the main loop, a dedicated unit computes $\hat{\bm{s}} = \bm{H}\,\hat{\bm{e}} \bmod 2$ for early-termination checking. This unit does not lie on the critical path but increases routing congestion and reduces the maximum achievable clock frequency; a folded architecture that reuses the CNU datapath for this computation could trade area for frequency in a future revision.

We do not process multiple check nodes concurrently within a single core: the BRAM capacity limits feasible parallelism, and concurrent check nodes would increase memory and data-dependency conflicts. A register-based architecture could reduce the prior-read latency but would require a large switching network to route the 397 messages to the CNU, lowering the maximum operating frequency without sufficient benefit.

\begin{figure}
    \includegraphics[width=\columnwidth]{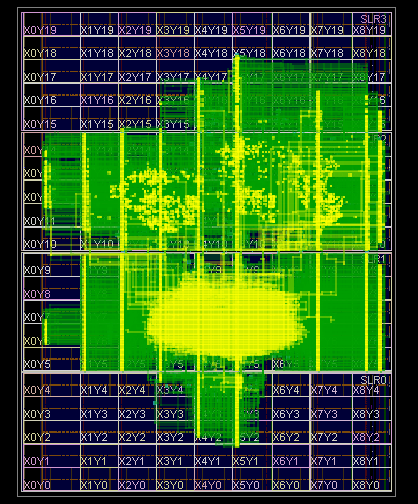}
    \caption{
        \textbf{Layout of a single core of the ensemble.}
    }
    \label{fig:layout_decoder}
\end{figure}
The implementation was synthesized with Vivado 2021.2 on the AMD xcvu19p-fsva3824-2-e device.
The corresponding FPGA layout is shown in Fig. \ref{fig:layout_decoder}.
The resource usage for one core comprises 98,088 LUT units, 5,983 LUTRAM units, 125,078 FF units and 1,145.50 BRAM units.
The total on-chip power is $14.2\,$W.
The clock period is 5.7~ns, yielding a per-iteration latency of $3.264\,$\textmu s. 
Because more than $50\,\%$ of the device's BRAM is occupied,
a second core cannot be fitted at the chosen word widths.
Reducing the number of bits per message or prior would save area at the cost of accuracy; we deliberately did not explore this trade-off in order to characterize the single-core decoder without additional quantization loss. The three-core ensemble therefore requires three separate FPGA devices.

For latency measurement, the core was exercised directly on the FPGA using an on-chip timer, without host communication.
An interface that sends the syndromes in parallel and receives the estimated errors from the decoder was tested using a Windows OS and a 1 Gb Ethernet connection, resulting in a total latency of 34$\,$\textmu s for the interface.
In a production system, optimized optical links or RTOS-based serial interfaces can reduce the communication overhead to hundreds of nanoseconds~\cite{liu_scalable_2026,amd_aurora_2024}. With such an interface the total decoding latency approaches the sub-microsecond average required by superconducting platforms at the lowest physical error rates, and remains more than two orders of magnitude below the 1~ms budget of neutral-atom and trapped-ion systems. A CPU-based decoder of equivalent accuracy would require many cores and a more complex interconnect to multiplex 250+ codes within 1~ms, and its timing jitter under a real-time operating system would be significantly larger than the deterministic FPGA latency.

\natcomm{
\section*{Data Availability}
Source data are provided with this paper.

\section*{Code Availability}
Code is available from the authors upon reasonable request.
}

\arxiv{
\section*{Code and Data Availability}
Code and source data are available from the authors upon reasonable request.
}

\natcomm{\bibliography{__main}}

\section*{Acknowledgments}
We thank Ioana Moflic for the suggestion to explore interleaved schedules.

\section*{Funding}
This work was supported by the R-QIP project within the DLR Quantum Computing Initiative (DLR QCI), funded by the Federal Ministry of Research, Technology and Space (BMFTR). Part of D.B.G. and F.G.H.’s work on the FPGA implementation and verification was supported by the project PID2023-147059OB-I00, funded by MCIU/AEI/10.13039/501100011033/FEDER, UE.

\section*{Author Contributions}
FL, DH and FGH conceived and supervised the study.
The measurement schedule was designed by AK with early investigation by DO and advice by AP, simulations have been performed by AK and the FPGA decoder was implemented and validated by DBG and FGH.
The paper was drafted by AK, FGH and FL.
It was discussed and revised by all authors.

\section*{Competing Interests}
The authors declare no competing interests.

\arxiv{\bibliography{__main}}


\end{document}